# MYSTERY OF THE "MAGIC NUMBER" 137:

# WAVE GENESIS, THEORETICAL REPRESENTATION,

# ROLE IN THE UNIVERSE

A. M. Chechelnitsky, Laboratory of Theoretical Physics,

Joint Institute for Nuclear Research,

141980 Dubna, Moscow Region, Russia

That mystery already exists more that 80 years.

Many of great theoreticans - founders of modern physics - Sommerfeld, Eddington, Born, Pauli, Dirac, Weyl, Heisenberg, Feynman, etc. deeply feel its true price, provocative, defiant character in connection with the fundamental basis of the theoretical phisics.

Is the Fine Structure Constant (FSC) the really fundamental constant or may be it varies with the time?

If that is the really fixed quantity, then - what is its true, theoretically based value?

Fundamental ideas of the Wave Universe concept open new possibilities to answer that and another incidentally arising questions.

## Challenge of Problem.

From all modern theories of microworld - quantum electrodynamics (QED) describes the dynamic structure and the interaction of elementary particles (photons, electrons, muons) most exactly.

There is the fundamental parameter (coupling constant, interaction parameter), that lies in the basis of that advanced and consistent theory - the Fine Structure Constant (FSC).

The theoretical representation of this constant is unknown up till now.

"The Mysterious Number 137" - so titled Max Born the famous paper[1] of 1936.

The Fine Structure Constant (FSC) $\alpha = 2\pi e^2/hc$ or nondimensional number $\alpha^{-1} \cong 137$ (where e - electron charge, h - Planck constant, c - speed of the light) was introduced in the theoretical physics by Arnold Sommerfeld[2] in 1915. That is the fundamental parameter of the all atomic spectroscopy. At present, only its experimental value is known.

Heizenberg[2]: "Various futile attempts were undertaken to derive the FSC; the most well-known - in the Eddington3 theory."

Born[2]: "That is the relation, for which the modern theory cannot propose the explanation, it is one of the most fundamental tasks of the future physics."

Sommerfeld[2]: "...And really: if Eddington is true, the conclusion that the elementary charge e may be constructed with using of quantum theory (h) and of theory relativity (c) must be inevitable. We believe in every miracle of quantum theory; we for a long time convinced, that the Planck constant h takes part in all elementary processes of non - animated nature.

But the construction of elementary charge perhaps may be its greatest triumph: it may discover incredible perspectives in simplifying the physical world picture. For confirming the tremendous significance, that the quantum theory, accordingly common opinion, has now, it is worth noting that if we still cannot speak confidently about such triumph, anyway we regard it as possible."

"...Value $\alpha = 2\pi e^2/hc = 1/137.008$ is the FSC (by Per Olyn). It is (apparently integer) numerical value so far remains one of the most mysterious puzzles of atomic physics."

Heisenberg[2]: "...Sommerfeld for a long time was convinced, that this constant after all is defined by laws of Nature, that it may be not an accidental parameter which, properly speaking, may accept any value."

Dirac[3]: "...Only about these nondimensional values we shall speak now. One of them is quantity reverse to the notorious fine structure constant - $hc/2\pi e^2$. It is a fundamental constant of atomic physics and approximately equals 137. Other nondimensional constant is difined by ratio of proton mass to electron mass $mp/me$ and is approximately 1840.

There does not exists satisfactory explanation of these numbers yet, but physicians hope that after all it will be obtained. Then these constants will be calculated with basic mathematical equations; it is quite possible that the similar constants consist of simple values like $4\pi$."

Weyl[5] "...The constant $\alpha$ (the square of e) is a clean nondimensional number which quals approximatly 1/137. The full theory must, by force of mathematical considerations, lead out this number as similar as geometry predicts, that the value of the number p (ratio of length of circle to its diameter) equals 3.1415... Whatever Eddington thinks about that, such theory does does not exist yet."

Feynman[6] : "...Nobody knows. It's one of the greatest damn mysteries of physics: a magic number that comes to us with no understanding by man. You might say the "Hand of God" wrote that number, and "we don't know how He pushed His pencil".

**Wave Universe.**

The understanding and interpretation of modern data about the properties of the surrounding orld, the structure of near space and deep space, about the structure of mega and micro systems which make up its hierarchy, are possible in the framework of the concept of a Wave Universe 7-15.

The notion of wave dynamic systems (WDS) is one of its central ideas.

An atom and the Solar system, molecules and galaxies are only some of their characteristic representatives in the micro and mega world which have essentially wave and megawave properties.

The existence in them of isolated stationary (elite) energy levels, corresponding spectrum of natural frequencies, and the presence of spectroscopy (megaspectroscopy) is a fundamental property of wave dynamic systems.

A broad spectrum of problems of not only astrophysics, but also sciences of the Earth, including, sciences of the biosphere, may be comprehended, within the framework of a Wave Cosmogeonomy - the science of the Cosmos and Earth, and their interaction.

**Wave Genesis of FSC.**

The Wave Universe conception and the Megaquantum Wave Astrodynamics (see 7-15) make it possible to understand (and calculate) the wave genesis of physically distinguished - planetary orbits, of nature rhythms, in particular, of Solar System rhythms-its megaspectroscopy - analog of the atom's wave spectroscopy.

It is for the first time that within the frames of Wave Universe conception we may naturally obtain the following surprisely simple analitical and numerical (closed) representation for the Fine Structure Constant that is proved to be correct by the logics of the consistant theory

**The Suggestion**

i) The Fine Structure Constant (FSC) equals

$$\alpha^{-1} = 239/4/2\pi = 137.0448088$$

ii) That is the fundamental constant not only of microworld (atoms), but also - of megaworld (astronomical systems) - one of the general nondimensional parameters of Universe.

The modern science dogma about existence of insuperable barrier in viewing the micro and megaworld objects dynamic structure may and must be exposed to the fundamental critical reconsideration.

**Data of Experiments and the Imperative of Theory.**

Forestolling the inevitably arising detailed critical discussion of situation connected with determination of Universe fundamental constants, we point out only to one fact then concerns to Fine Structure Constant.

The obtained theoretical value of FSC

$$\alpha^{-1} = 137.0448088$$

fully corresponds to data of the most precise Wilkinson and Crane's experiment, which is connected with the free electron magnetic momentum measurement [see Taylor,

Parker, Langenberg (1969), tabl. 28]16

$$\alpha^{-1} = 137.0467(36) \ (26 \cdot 10^{-4} \ \%)$$

Some observed difference between the theory result

$$\alpha^{-1} = 137.0448088$$

and the recommended now (Particle Data Group) worldwide, obtained by multitude set of different experiments value

$$\alpha^{-1} = 137.035989$$

demands special discussion.

Inevitably comes the time of critical analysis of the existing situation of experimental determination of fundamental constants (and FSC) and connected with its consequences approaches.

**REFERENCES**

1. Born, M. The Mysterious Number 137, Uspekhi Fis. Nauk, 16, N 6, p. 697, (1936) (in Russian).

2. Sommerfeld, A. Cognition Ways in Physics, Ed.by J.A. Smorodinsky, Nauka, Moscow, (1973) (in Russian).

3. Eddington, A.S. Fundamental Theory, Cambrdge University Press, (1948)

4. Dirac, P.A.M Directions in Physics, Ed. by H.Hora and Shepanski, John Wiley and Sons, N4, (1978).

5. Weyl, H. Selected Works, Nauka, Moscow, (1984) (in Russian, p.347)

6. Feynman, R.P. QED - The Strange Theory of Light and Matter, Alix G. Mautner Memorial Lectures Princeton University Press, Princeton, New Jersey, (1985)

7 Chechelnitsky A.M., Extremum, Stability, Resonance in Astrodynamics and Cosmonautics, M., Mashinostroyenie, 312 pp. (1980) (Monograph in Russian). (Monograph in Russian). (Library of Congress Control Number: 97121007; Name: Chechelnitskii A.M.).

8. Chechelnitsky, A.M., Megawave Genesis of the Solar System's Rhythms, Variation of Neutrino Flux and Cosmic Rays, Proceedings of the All-State Conference, in: "Muons and Neutrino Research in Large Water Volumes", Alma-Ata, KazGV, pp.44-52, (1983), (in Russian).

9. Chechelnitsky ,A.M., Wave Structure, Quantization, Megaspectroscopy of the Solar system; In the book:Spacecraft Dynamics and Space Research, M., Mashinostroyenie, pp.56-76 , (1986) (in Russian).

10. Chechelnitsky, A.M., Uranus System, Solar System and Wave Astrodynamics; Prognosis of Theory and Voyager-2 Observations, Doklady AN SSSR, v.303, N5 pp.1082-1088, (1988).

11. Chechelnitsky, A.M., Neptune - Unexpected and Predicted: Prognosis of Theory and Voyager-2 Observations, Report ( AF-92-0009 ) to the World Space Congress, Washington, DC, (Aug.22-Sept.5), (1992) .

12. Chechelnitsky, A.M., Wave Structure of the Solar System, Report to the World Space Congress, Washington, DC, (Aug.22-Sept.5), (1992).

13. Chechelnitsky A.M., Wave Structure of the Solar System, Tandem-Press, (1992) (Monograph in Russian).

14. Chechelnitsky A.M., Phythms Simphony of Cosmos, Earth, Biosphere - Megawave Genesis, Report to 30 COSPAR Assembly, Hamburg, 11-21 July (1994)

15. Chechelnitsky A.M., Magnetospheres and Heliosphere - As Phenomena of Wave Astrodynamics, Report to 30 COSPAR Assembly, Hamburg, 11-21 July (1994)

16. Taylor, B.N. Parker, W.H. Langenberg, D. N. The Fundamental Constants and Quantum Electrodynamics, Academic Press, New York, London, (1969)